\documentclass[aip,jap,reprint,floatfix,superscriptaddress,amsfonts,amsmath,amssymb]{revtex4-2}
\usepackage{graphicx}

\DeclareMathOperator{\Ei1}{E_1}
\begin{document}

\title{Effective medium theory for the electrical conductivity of random metallic nanowire networks} 

\author{Yuri Yu. Tarasevich}
\email[Corresponding author: ]{tarasevich@asu-edu.ru}
\affiliation{Laboratory of Mathematical Modeling, Astrakhan Tatishchev State University, Astrakhan, Russia}

\author{Irina V. Vodolazskaya}
\email{irina.vodolazskaya@asu-edu.ru}
\affiliation{Laboratory of Mathematical Modeling, Astrakhan Tatishchev State University, Astrakhan, Russia}

\author{Andrei V. Eserkepov}
\email{dantealigjery49@gmail.com}
\affiliation{Laboratory of Mathematical Modeling, Astrakhan Tatishchev State University, Astrakhan, Russia}

\date{\today}

\begin{abstract}
Interest in studying the conductive properties of networks made from randomly distributed nanowires is due to their numerous technological applications. Although the sheet resistance of such networks can be calculated directly, the calculations require many characteristics of the system (distributions of lengths, diameters and resistances of nanowires, distribution of junction resistance), the measurement of which is difficult. Furthermore, such calculations can hardly offer an analytical dependence of the sheet resistance on the basic physical parameters of the systems under consideration. Although various theoretical approaches offer such analytical dependencies, they are often based on more or less reasonable assumptions rather than rigorously proven statements. Here, we offer an approach based on Foster's theorem to reveal a dependence of the sheet resistance of dense nanowire networks on the main parameters of such networks. This theorem offers an additional perspective  on the effective medium theory and extends our insight. Since the application of Foster's theorem is particularly effective for regular random resistor networks, we propose a method for regularizing resistor networks corresponding to random nanowire networks. We found an analytical dependence of the effective electrical conductivity on the main parameters of the nanowire network (reduced number density of nanowires, nanowire resistance, and resistance of contacts between nanowires).
\end{abstract}

\maketitle

\section{Introduction\label{sec:intro}}
The effective medium theory (EMT) has been proposed by \citet{Bruggeman1935} to calculate physical constants of heterogeneous media such as dielectric constants and conductivities of composite media made of isotropic substances. EMT has as well been adapted to random resistor networks (RRNs), viz, regular networks with randomly distributed branch conductivities.\cite{Kirkpatrick1971,Kirkpatrick1973,Joy1978,Joy1979} For regular RRNs, the idea is to replace a network having random values of branch conductivity with a network of the same structure having identical values of all edge conductivities. In all those cases,\cite{Kirkpatrick1971,Kirkpatrick1973,Joy1978,Joy1979} derivations were based on the symmetry of the networks under consideration. However, symmetry is not a necessary requirement, the results are also valid for regular networks, i.e., networks, in which all nodes have the same valence (degree). The same results can be obtained on the basis of the Foster's theorem.\cite{Foster1961} \citet{Marchant1979} has demonstrated applications of the Foster's theorem to some particular resistor networks, viz., to regular networks having randomly distributed branch admittances, to irregular networks having identical branch admittances, and to irregular networks having randomly distributed branch admittances. Unfortunately, in the latter case, progress is rather modest as compared to the first two cases.
For a regular network ($\deg V = z$ for each node) with identical branch conductances $g = g_m$,
\begin{equation}\label{eq:MarchantRegular}
  \frac{g_m}{g_m + G'} = \frac{2}{z},
\end{equation}
where  $G'$ is the conductance between any two nearest nodes of the network when $g$ is disconnected.
Then, all $G'$ have the same value
\begin{equation}\label{eq:Marchanthomo}
  G' = g_m \left( \frac{z}{2} - 1 \right).
\end{equation}

The application of EMT to regular networks is equivalent to a replacement of the distribution of $g$, by a unique value $g_m$ obtained in an `effective network' of the \emph{same structure}, filled with identical conductances $g_m$ and given by~\eqref{eq:Marchanthomo}, which leads to the approximation\cite{Marchant1979} (cf.~\citet{Kirkpatrick1973})
\begin{equation}\label{eq:MarchantEMT}
  \int f_G(g) \frac{g_m - g}{g + g_m\left( \frac{z}{2} - 1  \right)} \, \mathrm{d}g \approx 0.
\end{equation}

Besides, \citet{Marchant1979}  explained the source of errors in the EMT predictions. \citet{Clerc1990} presented a review of works devoted to electrical conductivity of binary disordered systems.

Numerous technological applications of transparent conductive films (TCFs) based on randomly deposited metallic nanowires (random nanowire networks, RNNs) motivate interest in studying the properties of such random networks.\cite{Langley2013,Sannicolo2016,Papanastasiou2020,Ding2024}
Table~\ref{tab:AgNWs} presents average length, $\langle l_0 \rangle$, and diameter, $\langle d \rangle$,  of silver nanowires (AgNWs) used to produce TCFs. Deposited mass of AgNWs per unit area (areal mass density, amd) is also presented in Table~\ref{tab:AgNWs}. In modeling, it is customary to use the number density, i.e., the number of particles, $N$, per unit area,~$A$,
\begin{equation}\label{eq:n}
  n = \frac{N}{A}
\end{equation}
and it is convenient to take the length of the particle as the unit of length, i.e., use $nl_0^2$.
\begin{table}[!htb]
  \caption{Some characteristics of TCFs produced of AgNWs. }\label{tab:AgNWs}
  \centering
  \begin{ruledtabular}
  \begin{tabular}{lccc}
    Refs. & $\langle l_0 \rangle, \mu$m & $\langle d \rangle$, nm & amd, mg/m$^2$ \\
    \hline
  & 20.4 & 26 & 20--50  \\
     & 13.6 & 45 & 20--120  \\
    \citet{Lagrange2015} & 7.6 & 55 & 50--150 \\
     & 42.9 & 117 & 20--220 \\
     & 32.0 & 138 & 50--300 \\
    \hline
\citet{Bardet2021} & $8 \pm 3$ & $79 \pm 10$ & 25--250 \\
     & $11 \pm 5.8$ & $119 \pm 23$ &  \\
    \hline
    \citet{Sannicolo2016} & $\sim 10$ & 20--150 & 40--200 \\
    \citet{de2009} & 6.6 & 84 & 28--780 \\
    \citet{Celle2012} & 2--25 & 40--80 & 35 \\
    \citet{Nian2015} & 15 & 35 & 3.1--16 \\
    \citet{Goebelt2015} & 5.81 & 118 & 331 \\
    \citet{Yu2017} & 80 & 88 & 94--366.2 \\
    \citet{Milano2020} & 35 & 115 & 60--181 \\
    \citet{Suemori2020} & $10 \pm 5$ & $60 \pm $ 10& 23.7--39.1 \\
   \end{tabular}
  \end{ruledtabular}
\end{table}

Table~\ref{tab:AgNWs} suggests that the typical number density of nanowires $nl_0^2$ is few tens, while the maximal value in the exceptional cases may reach several hundreds. Obviously, that the minimal number density of nanowires has to correspond to the percolation threshold, since otherwise the system is insulating. The typical aspect ratio of nanowires, i.e., the ratio of nanowire length to its diameter is 100 of order of magnitude.

\citet{OCallaghan2016} studied the random nanowire networks (RNNs) accounting for both wire resistance and the contact resistance (so-called multinodal representation.\cite{GomesdaRocha2015}) They applied the EMT to find the dependency of the sheet resistance of transparent conductive films (TCFs) on the number density of nanowires. Several assumptions were stated and used in their consideration, viz, (i)~dense RNNs can be considered as square lattices with conductive edges, (ii)~optimal path between the two opposite border of the RNNs can be obtained using some ideas related to small-world networks, (iii)~lengths of wire segments correspond to Poisson distribution, (iv)~conductivities of contacts and wire segments are independently distributed.

Actually, the Poisson distribution describes the number of contacts per wire, while lengths of wire segments correspond to an exponential distribution.\cite{Yi2004} However,  \citet{OCallaghan2016} replaced the actual length distribution by average value of the segment length.

\citet{OCallaghan2016} demonstrated that, for both RNNs and a square lattice, dependencies of the two-point resistances on the distance between nodes are similar.  This observation was used to replace an RNN by a square lattice.\cite{OCallaghan2016}

\citet{Zeng2022} proposed another way how to correlate site percolation on a square lattice  to disordered nanowire networks (DNNs) and compared an EMT and an effective path theory (EPT).\cite{He2018}

Our goal is utilisation of the EMT with minimal and rigorously reasonable assumptions to find the dependence of the electrical conductivity of RNNs on the main physical parameters. The rest of the paper is constructed as follows. Section~\ref{sec:methods} provides necessary information and describes some technical details of simulation. Section~\ref{sec:results} presents the analytical approach, together with our main findings. Section~\ref{sec:concl} provides a discussion and summarizes the main results.

\section{Methods\label{sec:methods}}
Since the typical aspect ratio of nanowires is 100 of order of magnitude, the nanowires in simulations are often represented as zero-width conductive segments (sticks).\cite{Yi2004,Kumar2016,OCallaghan2016,Kim2018,Ainsworth2018} We use the same approach. Note that although model of zero-width sticks is widely used, it has some obvious limitations.  Since our focus are nanowires with large but finite aspect ratio, in our 2D model, the distance between the two nearest contacts has to be larger than the nanowire diameter (cf. \citet{Daniels2021}).

Let there be $N$ zero-width line segments (sticks) of equal length $l_0$. These sticks are deposited in such a way that coordinates of their centers are independent and identically distributed (i.i.d.) within the domain $L \times L$ with periodic boundary conditions (PBCs), while their orientations are equiprobable.
The electrical resistance of each stick is $R_w$, while the electrical resistance of each contact (junction) between segments is  $R_j$. In such a way, a RNN is mapped to a RRN.

Very important that obtained RRN is not a regular network despite $\langle z \rangle$ tends to 3 when RNNs under consideration are dense. We should emphasize that, in Ref.~\onlinecite{OCallaghan2016}, before EMT was applicable to RNNs, the real structure of the RRN under consideration was replaced by a square lattice. That replacement of the network structure rather than use a network of the same structure seems to contradict the basic idea of EMT.

\begin{figure*}
  \centering
  \includegraphics[width=\textwidth]{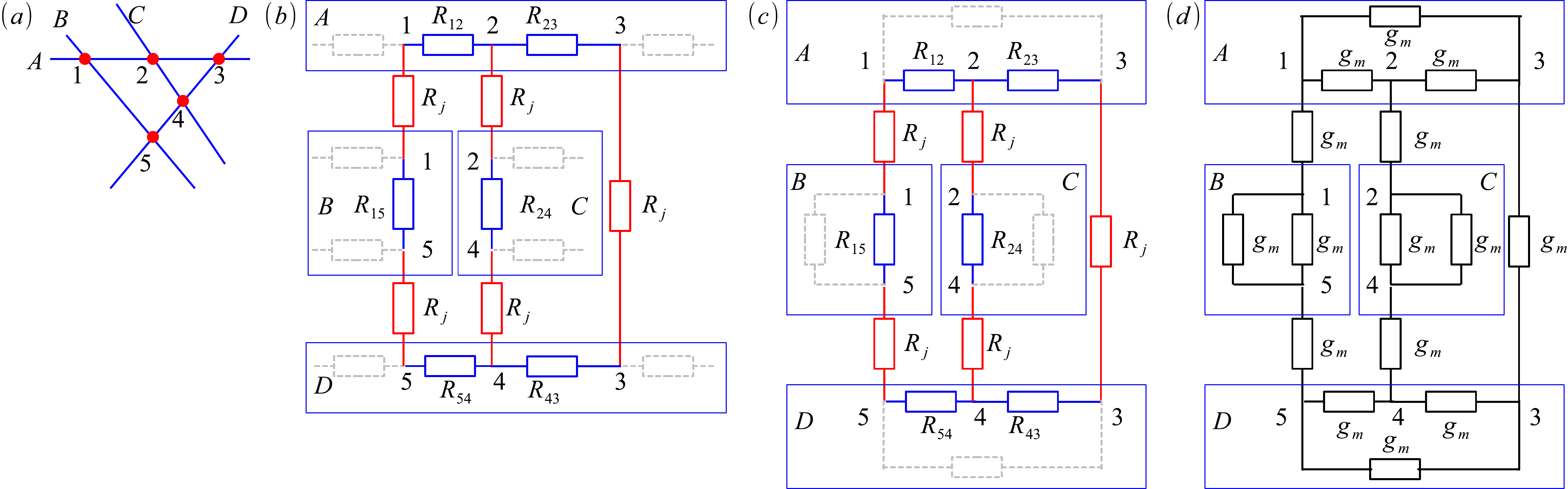}
  \caption{(a)~Example of a small nanowire network. (b)~Corresponding irregular random resistor network. (c)~An equivalent 3-regular random resistor network. (d)~An effective 3-regular resistor network with resistors of equal conductivity, $g_m$. }\label{fig:wires2resistors}
\end{figure*}

However, an RNN can be turned out into a regular network by applying a simple transformation. Let us mentally connect both ends of one nanowire. In order to preserve the electrical conductivity of the entire RNN, zero electrical conductivity should be assigned to this newly created edge connecting the extreme contacts on the nanowire, since the end segments of the nanowire do not participate in electrical conductivity. By applying such a closure to all nanowires, we obtain a 3-regular network, which has the same electrical conductivity as the original RNN (Fig.~\ref{fig:wires2resistors}).

Figure~\ref{fig:wires2resistors}a shows an example of a simple network formed by nanowires, which are marked by the letters $A, B, C, D$; the contacts between the nanowires are designated by numbers, $1, 2, \dots, 5$. Figure~\ref{fig:wires2resistors}b shows a resistor network corresponding to the original nanowire network. Resistors $R_{ij}$ correspond to the segments of nanowires between contacts $i$ and $j$; the end segments of nanowires that do not participate in the electrical conductivity are shown by dashed lines; resistors corresponding to segments of one nanowire are enclosed in a rectangle with the corresponding letter marking. Resistors $R_j$ correspond to contacts between nanowires. The resulting resistance network is irregular: it contains vertices of degree 3 and 1. Figure~\ref{fig:wires2resistors}c shows how the irregular resistor network presented in Fig.~\ref{fig:wires2resistors}b is regularized. Since current cannot flow through the resistors designated by a dotted line, zero conductivities can be assigned to them. If one interconnects two such resistors with zero conductivity, corresponding to the opposite ends of one nanowire, then, obviously, this will not affect the electrical conductivity of the entire network, but the network will become 3-regular, since all the vertices of the 1st order will disappear. Figure~\ref{fig:wires2resistors}d exhibits the final effective 3-regular resistor network with resistors of equal conductivity, $g_m$.

The average number of contacts (junctions) per a nanowire is
\begin{equation}\label{eq:kmean}
\langle k \rangle = \frac{2}{\pi} n l_0^2,
\end{equation}
(see, e.g., Refs. \onlinecite{Ainsworth2018,Heitz2011}). Associating a resistor, $R_j$, with each junction between the nanowires, we have
\begin{equation}\label{eq:N_j}
N_j = \frac{N\langle k \rangle }{2}=\frac{N n l_0^2 }{ \pi}
\end{equation}
such resistors.

In addition, there are $N$ resistor having zero conductivity, which correspond to the imaginary closed termini of nanowires,
\begin{equation}\label{eq:N0}
N_0 = N.
\end{equation}
Since there are on average $\langle k \rangle$ contacts on a nanowire, the average number of internal segments into which the nanowire is divided is $\langle k \rangle - 1$. The total number of internal segments in the entire system is
\begin{equation}\label{eq:Ns}
N_s = N\left(\langle k \rangle - 1\right).
\end{equation}
When the electrical resistance of a nanowire of length $l_0$ is $R_w$, then the resistance of a nanowire segment, $l$, between nearest junctions is $R_w l/l_0$. The corresponding number of such edges in the network is
\begin{equation}\label{eq:NumEdges}
N_s = N \left( \frac{2}{\pi} n  l_0^2 - 1\right).
\end{equation}
The total number of edges in the RNN is
\begin{equation}\label{eq:totalNumEdges}
N_E = N_j + N_0 + N_S 
=\frac{3}{2}\langle k \rangle N.
\end{equation}
A fraction of edges having the resistance  $R_j$ is
\begin{equation}\label{eq:fracJ}
\frac{N_j}{N_E} = \frac{N\langle k \rangle }{ 2} \frac{2}{3\langle k \rangle N} = \frac{1}{3}.
\end{equation}
A fraction of edges having the resistance  $R_w l/l_0$ is
\begin{equation}\label{eq:FracIntEdges}
\frac{N_s}{N_E} = \frac{ N\left(\langle k \rangle - 1\right)}{\frac{3}{2}\langle k \rangle N}= \frac{2} {3}\left(1 - \frac{1}{\langle k \rangle }\right) = \frac{2} {3} - \frac{\pi}{3n  l_0^2} .
\end{equation}
A fraction of edges having the infinite large resistance  is
\begin{equation}\label{eq:FracInf}
\frac{N_0}{N_E} = \frac{N}{\frac{3}{2}\langle k \rangle N}=\frac{2}{3\langle k \rangle} = \frac{\pi}{ 3 n  l_0^2}.
\end{equation}
The conductance PDF in the RNN is
\begin{multline}\label{eq:PDFRNN}
f_G(g)\\ = \frac{1 }{3} f^{(1)}_G(g)  + \frac{\pi}{ 3 n  l_0^2} f^{(2)}_G(g)  + \left(  \frac{2} {3} - \frac{\pi}{3n  l_0^2} \right) f^{(3)}_G(g),
\end{multline}
where $f^{(1)}_G(g) =\delta \left(g - R_j^{-1}\right)$
is the PDF of edges, corresponding to junctions, $ f^{(2)}_G(g)  =  \delta(g)$
is the PDF of imaginary nanowire segments connecting end junctions at the nanowire.
Here, $\delta(x)$ means the Dirac delta function.  $ f^{(3)}_G(g) $ is the PDF of nanowire segments between nearest junctions. The conductance of any edge (segment) is inversely proportional to its length, viz., $ g(l) = l_0/(l R_w)$, where $l$ is the segment length.

Since, when the number density of sticks is large enough, the number of contacts on a stick obeys a Poisson distribution, lengths of segments between two nearest contacts are expected to obey an exponential distribution with the mean $\langle l \rangle $~\cite{Yi2004}
 \begin{equation}\label{eq:PDFapprox}
 f_L(l;\langle l \rangle ) = \frac{1}{\langle l \rangle} \exp\left( -\frac{l}{\langle l \rangle }\right),
\end{equation}
where
 \begin{equation}\label{eq:lmean}
\langle  l  \rangle = \frac{ l_0}{\langle k \rangle} =   \frac{\pi }{2 n  l_0}.
\end{equation}
Then, $f^{(3)}_G(g)$ can be deduced from \eqref{eq:PDFapprox} as follows
\begin{equation}\label{eq:PDFg}
  f^{(3)}_G(g)  = \frac{l_0}{R_w g^2} f_L\left( \frac{l_0}{R_w g} ; \langle l \rangle \right).
\end{equation}
Hence,
\begin{equation}\label{eq:PDFg3}
  f^{(3)}_G(g)  =  \frac{2 n  l^2_0}{\pi R_w g^2} \exp\left( -\frac{2 n  l^2_0}{\pi R_w g} \right).
\end{equation}

\section{Results\label{sec:results}}

\subsection{Both junction resistances and wire resistance are accounted for\label{subsec:JWR}}
Since the effective network under consideration is a 3-regular network with randomly distributed resistances of edges, hence, according to Ref.~\onlinecite{Marchant1979}, \eqref{eq:MarchantEMT} is as well valid in this case, if PDF $f_G(g)$ corresponds the above described distribution
\begin{multline}\label{eq:PDFJWR}
f_G(g) =  \frac{1 }{3}\delta \left(g - R_j^{-1}\right) + \frac{\pi}{ 3 n  l_0^2} \delta(g)\\
+ \left( \frac{2} {3} - \frac{\pi}{3n  l_0^2} \right)\frac{2 n  l^2_0}{\pi R_w g^2} \exp\left( -\frac{2 n  l^2_0}{\pi R_w g} \right).
\end{multline}

Since in our case  $\deg V = 3$,
\begin{equation}\label{eq:MarchantRegular1}
G' = \frac{g_m}{2}
\end{equation}
(cf. \citet{Joy1979,Tarasevich2024}).
For a regular network with different branch conductances
\begin{equation}\label{eq:MarchantRegular2}
  \left\langle\frac{g}{g + G'}\right\rangle = \frac{2}{ \deg V  }.
\end{equation}
Then, the effective medium conductance, $g_m$, can be found by subtracting \eqref{eq:MarchantRegular} and \eqref{eq:MarchantRegular2}
\begin{equation}\label{eq:EMTint}
\int\limits_{g_0}^\infty f_G(g) \frac{g_m - g}{g_m/2 + g} \, dg=0.
\end{equation}
Then,
\begin{multline}\label{eq:MarchantRegular3}
\left(2 - \frac{\pi}{n  l _0^2} \right)\int\limits_0^\infty \frac{2 n  l^2_0}{\pi R_w g^2} \exp\left( -\frac{2 n  l^2_0}{\pi R_w g} \right) \frac{g_m - g}{g_m + 2 g} \,dg\\
+ \frac{g_m R_j - 1}{g_m R_j + 2 }  + \frac{\pi}{ n  l _0^2}  = 0.
\end{multline}
Here, we extended an integration limit, accounting for that the PDF decreasing very fast when the number density of wires is large.
\begin{multline}\label{eq:MarchantRegular4}
\left(2 - \frac{\pi}{n  l _0^2} \right)\int\limits_0^\infty \frac{2 n  l^2_0}{\pi R_w g} \exp\left( -\frac{2 n  l^2_0}{\pi R_w g} \right) \frac{ dg}{g_m + 2 g} \\
=\frac{g_m R_j + 1}{g_m R_j + 2 }.
\end{multline}
Using the exponential integral, $\Ei1(x)$, we get
\begin{multline}\label{eq:solution}
2\frac{2 n  l^2_0}{\pi R_w g_m}\mathrm{Ei}_{1} \left(2\frac{2 n  l^2_0}{\pi R_w g_m}\right) \exp\left(2\frac{2 n  l^2_0}{\pi R_w g_m}\right)\\
=\frac{g_m R_j + 1}{g_m R_j + 2 }\left(1 - \frac{\pi}{2 n  l _0^2} \right)^{-1}.
\end{multline}
When $R_j =0$, numerical solution of \eqref{eq:solution} gives
\begin{equation}\label{eq:gmWDR}
g_m R_w \approx 3.275 \frac{2 n  l^2_0}{\pi },
\end{equation}
hence, $g_m R_w = 2.08 n l_0^2$.

\begin{figure}[!htb]
  \centering
  \includegraphics[width=\columnwidth]{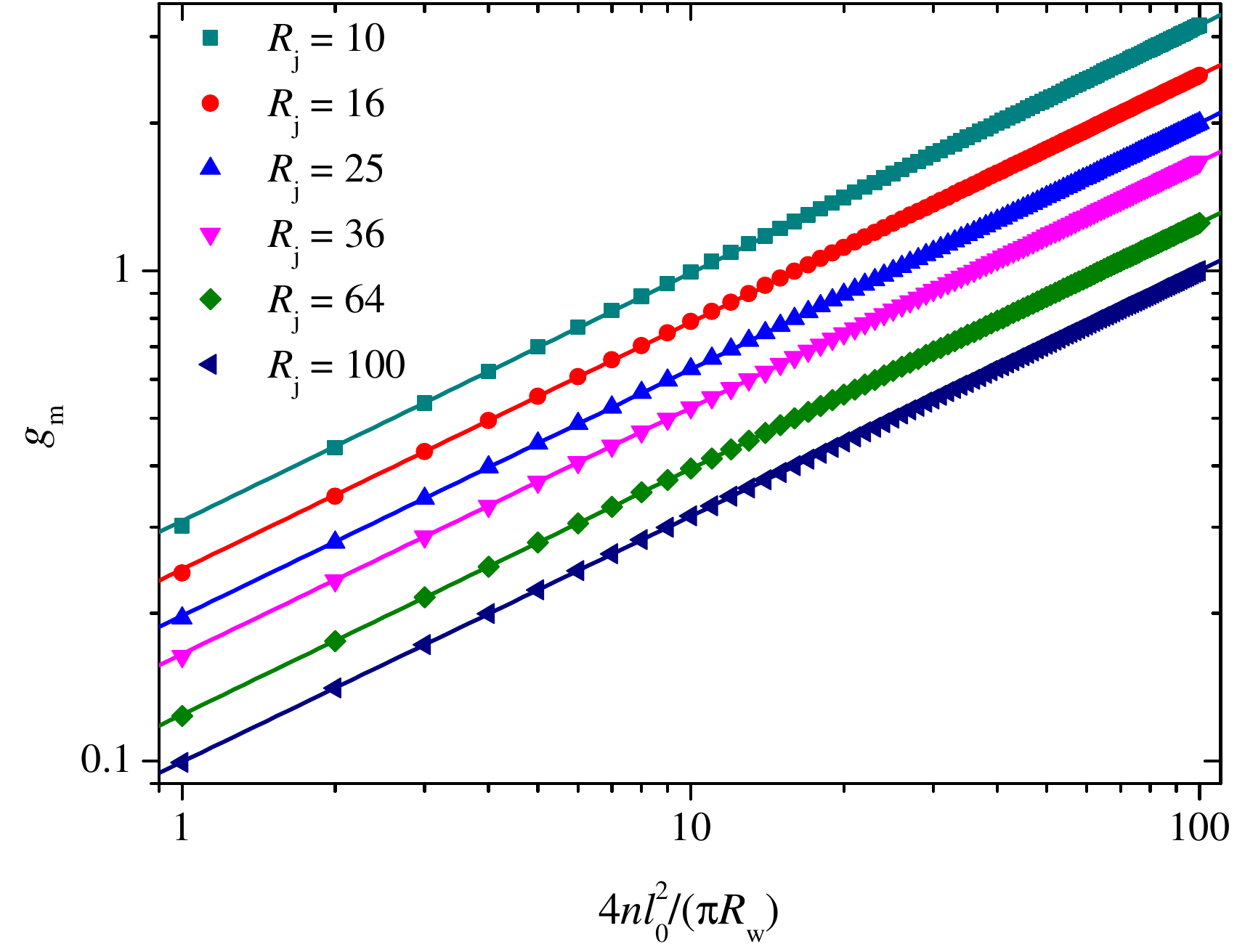}
  \caption{Dependence of the effective conductivity, $g_m$, on the effective number density of nanowires for different values of the junction resistance, $R_j$. Lines correspond to the least squares fits.}\label{fig:gm}
\end{figure}

In the common case, \eqref{eq:solution} was found for different values of $R_j$. Figure \ref{fig:gm} presents dependence of the effective conductivity, $g_m$, on the quantity $4 n  l^2_0/(\pi R_w)$.

Table \ref{tab:LSF} presents results of the least squares fitting
\begin{equation}\label{eq:LSFJWR}
\log_{10} g_m = A \log_{10} \frac{4 n l^2_0}{\pi R_w} + B.
\end{equation}
These results suggest that, for large values of $R_j$, the effective conductive can be written as follows
\begin{equation}\label{eq:gmJWR}
g_m = \sqrt{\frac{4 n l^2_0}{\pi R_j R_w}}.
\end{equation}
\begin{table}[!htb]
  \centering
  \caption{Least squares fit}\label{tab:LSF}
  \begin{ruledtabular}
  \begin{tabular}{llll}
  $R_j$ & $A$ & $B$ & $10^{-2B}$ \\
    \hline
1 & 0.512 & -0.026 & 1.13 \\
2 & 0.510 & -0.171 & 2.2 \\
10 & 0.505 & -0.510 & 10.5 \\
16 & 0.504 & -0.609 & 16.5 \\
25 & 0.503 & -0.704 & 25.6 \\
36 & 0.502 & -0.782 & 36.6\\
64 & 0.501 & -0.905 & 64.6 \\
100 & 0.501 & -1.001 & 100.6\\
      \end{tabular}
      \end{ruledtabular}
\end{table}

An assumption that the wire resistance is negligible as compared to the junction resistance is widely used.\cite{GomesdaRocha2015,Kim2019,Jagota2020,AlvarezAlvarez2021} This assumptions is quite natural for dense nanowire networks. When $R_w=0$, $f^{(3)}_G(g)  = 0$, thus, \eqref{eq:MarchantRegular3} turns out in
\begin{equation}\label{eq:JDR}
\frac{g_m R_j - 1}{g_m R_j + 2 }  + \frac{\pi}{ n  l _0^2}  = 0.
\end{equation}
Solution of this equation is
\begin{equation}\label{eq:gmJDR}
g_m R_j =\frac{n l_0^2 - 2\pi}{n l_0^2 + \pi}.
\end{equation}

\subsection{Seamless networks}
When only the wire resistances are accounted for, the RNN is a planar network, which tends to a 4-regular network  as the number density of wires increases, while the resistances of edges are random variables which values are proportional to edge lengths \eqref{eq:PDFapprox}. Again, connecting both dead-ends of each nanowire, we got a 4-regular network, in which $N$ resistors have infinite large resistance, while $N(\frac{2}{\pi} n l_0^2 - 1)$ resistors have the resistance $R_w l/l_0$.
\begin{multline}\label{eq:MarchantEMT1}
    \left(1-\frac{\pi}{2nl^2_0} \right)\int\limits_{0}^\infty\frac{2 n  l^2_0}{\pi R_w g^2} \exp\left( -\frac{2 n  l^2_0}{\pi R_w g} \right) \frac{g_m - g}{g_m + g} \, \mathrm{d}g\\
    +\frac{\pi}{2nl^2_0}  \int\limits_{0}^\infty\delta(g)  \frac{g_m - g}{g_m + g} \, \mathrm{d}g = 0.
\end{multline}
Integration leads to the following equation
\begin{multline}\label{eq:seamless}
 \frac{2 n  l^2_0}{\pi R_w g_m} \Ei1\left(\frac{2 n  l^2_0}{\pi R_w g_m}\right) \exp\left(\frac{2 n  l^2_0}{\pi R_w g_m}\right)\\= \frac{1}{2}\left(1-\frac{\pi}{2nl^2_0} \right)^{-1},
\end{multline}
which solution is
\begin{equation}\label{eq:WDRroot}
g_m R_w  \approx 1.637\frac{2n l^2_0  }{\pi },
\end{equation}
i.e., $g_m R_w \approx 1.04 n l^2_0 .$

Note that the obtained result is valid for \emph{any} 4-regular infinite graph, e.g., square lattice, kagome lattice,  rhombitrihexagonal tiling of the Euclidean plane (3.4.6.4),  etc. Although all these examples are related to planar graphs, planarity is not a necessary requirement. However, for any particular 4-regular graph, a dependency of the sheet resistance on the edge conductance is individual. Thus, for the square lattice, $R_\Box = g_m^{-1}$, while, for the kagome lattice, $R_\Box = 2(g_m\sqrt{3})^{-1}$. The problem is, the network under consideration is a random graph; although its edge conductance is now known, we cannot still find its sheet resistance.

\begin{figure}[!hbt]
  \centering
  \includegraphics[width=\columnwidth]{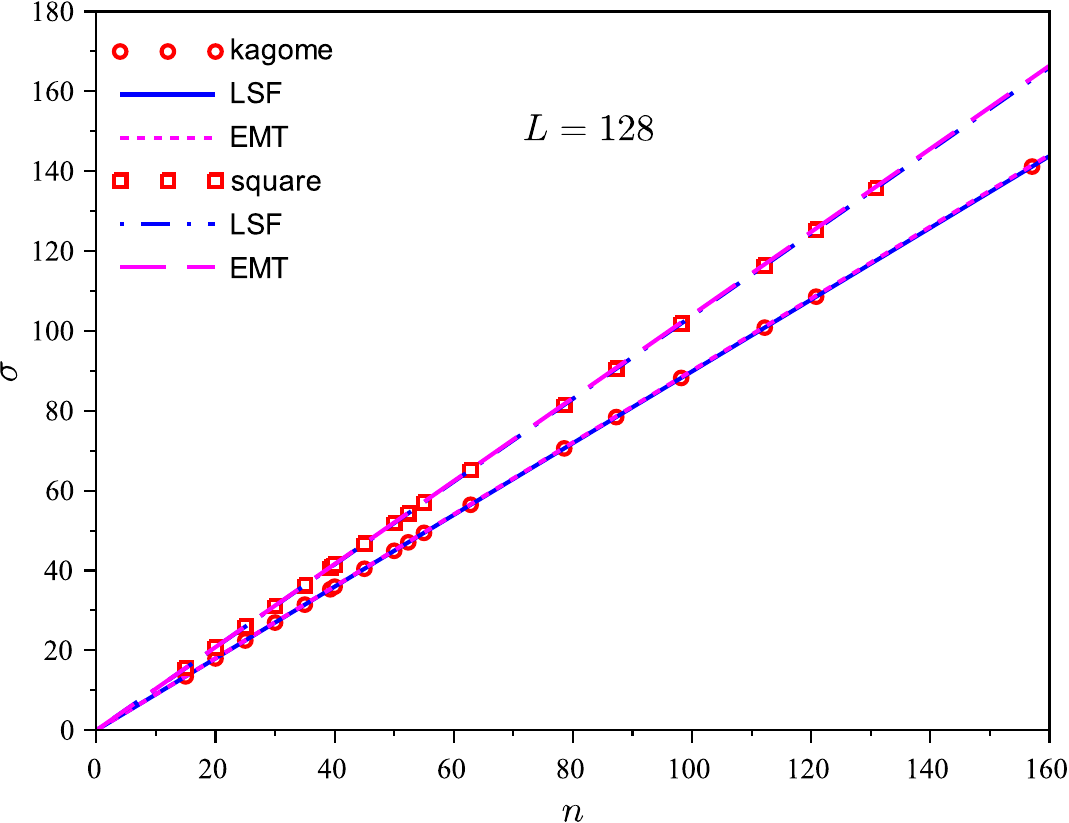}
  \caption{Dependencies of the electrical conductance on the number density edges for a square lattice and a kagome lattice of linear size $L=128$ and randomly distributed values of the edge conductances according to~\eqref{eq:PDFg}. Results are averaged over 10 independent runs.}\label{fig:squareEMT}
\end{figure}

To check the obtained value of the effective conductance, we used a computational test. A square lattice and a kagome lattice are instances of a 4-regular networks. In a bus-bar geometry, we calculated the conductivities of these two lattices ($L=32, 64,$ and $128$) with random values of the edge conductances according to PDF~\eqref{eq:PDFg} with $g_0=1$, $l_0 =1$. The results were averaged over 10 independent runs for each value of $\langle l \rangle $. Bearing in mind RNWs, we presented in Fig.~\ref{fig:squareEMT} the electrical conductance of the lattices against the number density of edges rather than $\langle l \rangle $. The least squares fit (LSF) suggests that the conductance is proportional to $1.037n$ in the case of a square lattice. Since in a bus-bar geometry, the resistance of a square resistor network $L \times L$ equals a resistance of one branch, in fact, Fig.~\ref{fig:squareEMT} evidenced that the electrical conductance of random square lattice is equal to the electrical conductance of square lattice with identical conductivity of each edge 1.04. In the case of a kagome lattice, LSF evidenced that the slope is 0.899, while the theoretical prediction suggests $0.901$. We consider this as a confirmation of obtained value the effective conductance $g_m$.

\begin{figure}[!hbt]
  \centering
  \includegraphics[width=\columnwidth]{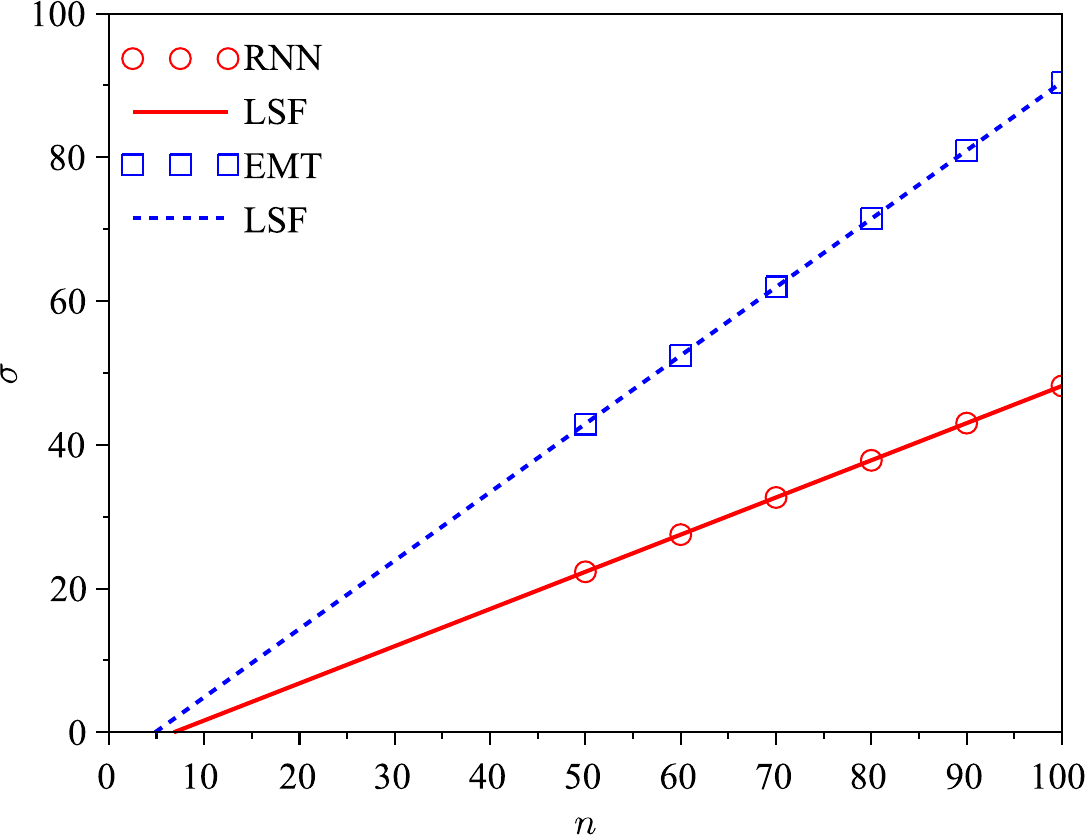}
  \caption{Dependencies of the electrical conductance on the number density of wires for a RNN of linear size $L=32$. Results are averaged over 10 independent runs.}\label{fig:RNNEMT}
\end{figure}

Figure~\ref{fig:RNNEMT} compares the electrical conductivity of RNNs obtained using the direct computations (circles) and EMT (squares). In the latter case, all edge conductivities were set according to obtained value of $g_m$. The EMT gives about twice larger conductivity as compared to direct computations. Since EMT have demonstrated an excellent agreement for regular lattices (square and kagome), we can see only one reason of the discrepancy in the case of RRN. While square lattice and kagome lattice are \emph{strictly} 4-regular, the RRN is \emph{almost} 4-regular, i.e., dead-ends are crucial. In other words, the EMT should be applied only to the equivalent RRNs which correspond to RNNs with closed termini of nanowires.

\section{Discussion\label{sec:concl}}
When the network structure and the resistances of each wire segment and each contact between wires are known, the electrical resistance of the random nanowire network can be calculated directly, however, analytical dependence of the resistance on physical and geometric parameters are difficult to extract from these calculations, while measuring all the quantities necessary for calculations is an extremely complicated task.\cite{Kim2018,Kim2019} In contrast, theoretical approaches supply us with analytical dependencies of the sheet resistance on the main parameters for a class of networks if statistical properties of network structure and of the main physical parameters are known. Different variants of the mean-field approach have been utilized to obtain the dependencies of the effective resistance on the main parameters.\cite{Kumar2017,Forro2018,Tarasevich2022} Alternatively, utilization of a probabilistic graph Laplacian was proposed to obtain the electrical conductivity of the random nanowire networks when only the junction resistance are accounted for.\cite{Jagota2020} However, analytical approaches overestimate the electrical conductivity as compared to direct computations. It appears that all possibilities for further improvement of conductivity estimates using the mean-field approach have been completely exhausted.\cite{Tarasevich2022,Tarasevich2023}

The EMT has been successfully applied to determine the effective resistance of random resistor networks having \emph{regular structure and translational symmetry}.\cite{Kirkpatrick1973,Joy1978,Joy1979} To our best knowledge, there are only two examples where the EMT was utilized to find the resistance of random nanowire networks, i.e. \emph{irregular networks}.\cite{OCallaghan2016,Zeng2022} In both cases, the real network structure was replaced by a square lattice. Such a replacement does not seem convincing and justified. In particular, similar dependencies of the two-point resistance on the distance between nodes for a square network and a random nanowire network were used as a justification for replacing the real random nanowire network having an irregular structure with a square lattice.\cite{OCallaghan2016} In our opinion, such similarity only indicates that in both cases the systems are two-dimensional, so the logarithmic dependence of the two-point resistance on the distance is quite expected.\cite{Tarasevich2025}
On the contrary, the network regularization method we proposed is justified and simple.

Using Foster's theorem along with rigorously reasonable assumptions, we found the dependence of the effective electrical conductivity of the edges of random nanowire networks on the main physical parameters. We demonstrated that a direct application of the effective medium theory to the random nanowire networks leads to a significant overestimate of the electrical conductance since such networks are not regular networks. A regularized random resistor network should be used instead.

Although for simplicity we have considered systems in which all wires have the same length and their orientations are equally probable, the generalization is obvious. Indeed, the basic formulas can be written not only using the number density of nanowires, $n$, but also using the average number of contacts per wire, $\langle k \rangle$. The latter quantity is expressed in terms of the probability of two wires crossing, $P$, viz.,
$$
\langle k \rangle = P N, \text{ when } N \gg 1.
$$
The probability that a wire of length $l_2$ intersects a wire of length $l_2$ is
\begin{widetext}
$$
P = \int\limits_{l_\text{min}}^{l_\text{max}}\int\limits_{l_\text{min}}^{l_\text{max}}\int\limits_{-\pi/2}^{\pi/2}\int\limits_{-\pi/2}^{\pi/2}\frac{l_1 l_2}{A} \sin |\theta_1 - \theta_2| f_L(l_1) f_L(l_2) f_\Theta(\theta_1)f_\Theta(\theta_2) \,dl_1\,dl_2\,d\theta_1\,d\theta_2,
$$
\end{widetext}
where $f_L(l)$ is the wire length PDF and $f_\Theta(\theta)$ is the wire orientation PDF. Since length and angular distributions are independent,
$$
P = C\frac{\langle l \rangle^2 }{A},
$$
where
$$
C = \int_{-\pi/2}^{\pi/2}  \int_{-\pi/2}^{\pi/2}\sin |\theta_1 - \theta_2| f_\Theta(\theta_1) f_\Theta(\theta_2)  \,d\theta_1 \,d\theta_2.
$$
For isotropic systems, $C = 2/\pi$.

Thus, in the common case,
$$
\langle k \rangle = C n \langle l \rangle^2.
$$
This means that, in the common case, the effective conductivity can be obtained in the same way but using this formula for $\langle k \rangle $.

The proposed here method of regularization can be applied to random networks of (nano)wires of any number density. However, the distribution of conductor segment lengths was obtained assuming that the number of wires is large. In addition, our method of regularization is quite applicable in the case when the (nano)wires are curved or wavy. In this case, the calculation of the probability of wires intersection can be carried out based on work \citet{Yi2004}.

Although the EMT allows to find edge conductances to replace a regular random resistor network by a resistor network of the same structure and identical resistances of all its branches, this cannot assist in finding of the sheet resistance unless the case networks having a translational symmetry, e.g., square, honeycomb, kagome, etc. Otherwise, different tricks should be used to map the network under consideration to a square lattice.\cite{OCallaghan2016,He2018,Zeng2022} These tricks can hardly be considered as well founded.
Thus, even consistent application of the EMT to random nanowire networks does not remove the main problem, viz., although the value of the effective electrical conductivity of network edges is now known, this does not help to find the effective conductivity of the entire network, since the dependence of the effective electrical conductivity on the edge resistance depends significantly on the structure of the particular network.

\citet{Jagota2020} proposed an approximate analytical model for sheet conductivity of random nanowire networks. They accounted only for junction resistances and used an averaged Laplace matrix (probabilistic graph Laplacian) to find the electrical conductivity. However, the proposed approach seems to be applicable to any regular network with identical branch resistances, i.e., to those networks which are obtained after the replacement of all branch resistances to~$g_m$. If an information about spectral properties of the probabilistic graph Laplacian can be obtained, application of Wu's theorem\cite{Wu2004} could supply us with the sheet resistance.

\section*{Data Availability Statement}
Data available on request from the authors.

\begin{acknowledgments}
Y.Y.T. acknowledges partial funding from the FAPERJ, Grants No.~E-26/202.666/2023 and No.~E-26/210.303/2023 during his stay in Instituto de F\'{\i}sica, Universidade Federal Fluminense, Niter\'{o}i, RJ, Brasil. Authors are thankful to Avik Chatterjee for careful reading of the manuscript and discussions.
\end{acknowledgments}

\bibliography{EMT2025}

\end{document}